\documentclass[12pt,a4paper,oneside]{article}
\usepackage[authoryear]{natbib}
\usepackage{times,epsfig,makeidx,amsfonts,amsmath,dcolumn,enumerate,multirow,graphicx,amssymb,colortbl,bm}
\usepackage[margin=1in]{geometry}
\usepackage{algorithm}
\usepackage[]{algorithmic}
\usepackage{epstopdf}
\usepackage{authblk}
\usepackage{afterpage,lscape}
\usepackage{breakcites}

\newtheorem{definition}{Definition}

\newcommand*\colvec[3][]{
    \begin{pmatrix}\ifx\relax#1\relax\else#1\\\fi#2\\#3\end{pmatrix}
}

\begin{document}
\title{On modelling asymmetric data using two--piece sinh-arcsinh distributions}
\author[*]{{Francisco J. Rubio}\footnote{E-mail:  Francisco.Rubio@warwick.ac.uk}}
\author[**]{{Emmanuel O. Ogundimu}\footnote{E-mail:  emmanuel.ogundimu@ndorms.ox.ac.uk}}
\author[*]{{Jane L. Hutton}\footnote{E-mail:  J.L.Hutton@warwick.ac.uk}}
\affil[*]{Department of Statistics, University of Warwick}
\affil[**]{Centre for Statistics in Medicine, University of Oxford}

\renewcommand\Authands{ and }
\maketitle

\begin{abstract}
We introduce the univariate two--piece sinh--arcsinh distribution, which contains two shape parameters that separately control skewness and kurtosis. We show that this new model can capture higher levels of asymmetry than the original sinh--arcsinh distribution \citep{JP09}, in terms of some asymmetry measures, while keeping flexibility of the tails and tractability. We illustrate the performance of the proposed model with real data, and compare it to appropriate alternatives. Although we focus on the study of the univariate versions of the proposed distributions, we point out some multivariate extensions.
\end{abstract}

\noindent {\it Key Words: Interpretability of the parameters; kurtosis; skewness; skew--symmetric.}

\section{Introduction}

Univariate parametric flexible distributions that can capture departures from normality in terms of asymmetry and kurtosis have been widely studied. This interest is often motivated by the fact that these distributions can produce robust models. Flexible distributions are typically, but not exclusively, obtained by adding parameters to a symmetric distribution. These methods can be classified either as parametric transformations of a distribution function \citep{FS06} or as parametric changes of variable \citep{LP10a}. We do not provide an extensive overview of the literature on these classes, but only present a brief summary of the methods that are relevant to this work. We refer the reader to \cite{J14} for a good survey of flexible distributions. One of the most popular distributions obtained as a transformation of a symmetric distribution is the \emph{skew normal} (SN) proposed by \cite{A85}. Its construction consists of multiplying the normal density by a parametric \emph{skewing function}, as follows:

\begin{eqnarray}\label{skewnormal}
g(x;\lambda)= 2\phi(x)\Phi(\lambda x),
\end{eqnarray}

\noindent where $\lambda \in{\mathbb R}$, $\phi$ and $\Phi$ denote the standard normal density and distribution function, respectively. It is easy to see that density (\ref{skewnormal}) is asymmetric for $\lambda\neq0$ and converges to the right/left half--normal as $\lambda \rightarrow \pm \infty$. \cite{WBG04} showed that this idea can be extended to any symmetric probability density function (pdf) $f$ with support on ${\mathbb R}$ through the transformation:

\[ g(x;\lambda)= 2f(x)\pi(\lambda x),  \]

\noindent where $\pi$ is a nonnegative function satisfying $\pi(x)+\pi(-x)=1$. The distributions obtained with this technique are usually referred to as \emph{skew--symmetric models}. Although this method leads to a tractable expression for the density function, some skew--symmetric models have inferential problems. For instance, \cite{A85} showed that the Fisher information matrix of the SN distribution is singular when the skewness parameter $\lambda$ is zero, which also leads to the presence of flat ridges in the likelihood surface \citep{P00}. Another strategy for adding shape parameters to a distribution $F(x)$ consists of raising this function to a positive power $\alpha$, leading to the class of power distributions. We refer the reader to \cite{P12} for a survey of the properties of this transformation as well as some inferential properties. Another popular method is the \emph{two--piece} transformation \citep{F97,FS98,MH00,AValle05,J06}, which consists of using different scale parameters on either side of the mode of the density under several parameterisations. Although standard likelihood theory is not applicable in the family of two--piece distributions, due to the lack of differentiability (of second order) of the corresponding density function at the mode, it has been shown that maximum likelihood (ML) estimation is well--behaved \citep{JA10}, especially under certain parameterisations that induce parameter orthogonality. Further, some asymptotic results have been proven for the maximum likelihood estimators of the parameters of some of these distributions \citep{MH00,AValle05,ZG10,JA10}. The sinh--arcsinh (SAS) distribution \citep{JP09} represents an interesting model obtained as a parametric change of variable. This distribution, which is described in the next section, contains two shape parameters that can be interpreted as skewness and kurtosis parameters, and has tractable expressions for the density and distribution functions. Another appealing property is that it contains models with both heavier or lighter tails than those of the normal distribution. However, we will show in the next section that this model cannot accommodate high levels of skewness in terms of some interpretable measures of asymmetry.

We propose a flexible distribution obtained by applying the two--piece transformation to the symmetric sinh--arcsinh distribution, which we call the two--piece sinh--arcsinh distribution (TP SAS). The reader may naturally question the need for another model and the value of this approach to modelling asymmetry. Our justification is modest but still valid: we try to produce a distribution that can capture higher levels of skewness than the original SAS distribution while keeping the tail flexibility, ease of use, and appealing inferential properties. We also argue in favour of the proposed distribution using the interpretability of its parameters. Concerning the value of this approach, we compare it to the skew--symmetric extension of the symmetric SAS distribution. The resulting distribution, denoted SS SAS, can also capture higher levels of asymmetry than the SAS distribution, but also inherits the inferential issues of the skew normal distribution, which is a particular case of the SS SAS. This raises another question: which of the three versions of the sinh--arcsinh distribution (SAS, TP SAS, SS SAS) should we use?  There are, of course, many formal model selection tools for use in applications. However, we argue that other features such as ease of use, inferential properties, and interpretability of the model parameters have to be considered as well, especially in cases when the model selection tools do not clearly favour one of the competitors.

The paper is organised as follows. In Section \ref{OSAS} we provide a brief summary of the SAS distribution. We also study the flexibility of this distribution in terms of some measures of skewness. In Section \ref{TPSAS} we introduce the TP SAS and SS SAS distributions and discuss some basic distributional properties. The performance of the proposed models is illustrated with an example in Section \ref{Examples}.

\section{The original sinh--arcsinh distribution}\label{OSAS}

The SAS cumulative distribution function (cdf) \citep{JP09} is obtained by applying the parametric change of variable $H(x;\mu,\sigma,\varepsilon,\delta) = \\ \sinh \left( \delta \operatorname{arcsinh} \left( \dfrac{x-\mu}{\sigma} \right) - \varepsilon\right)$ to a normal random variable, as follows:

\begin{eqnarray}\label{SASCDF}
S_0(x;\mu,\sigma,\varepsilon,\delta)= \Phi\left[H(x;\mu,\sigma,\varepsilon,\delta)\right],
\end{eqnarray}

\noindent where $x\in{\mathbb R}$, $\mu \in{\mathbb R}$ is the location parameter, $\sigma\in{\mathbb R}_+$ is the scale parameter, $\varepsilon\in{\mathbb R}$, and $\delta\in{\mathbb R}_+$. The corresponding density function can be obtained in closed form by differentiating (\ref{SASCDF}) as follows:

\begin{eqnarray}\label{SASPDF}
s_0(x;\mu,\sigma,\varepsilon,\delta)= \phi\left[H(x;\mu,\sigma,\varepsilon,\delta)\right] h(x;\mu,\sigma,\varepsilon,\delta),
\end{eqnarray}

\noindent where $h(x;\mu,\sigma,\varepsilon,\delta) = \dfrac{\delta \cosh\left(\delta\operatorname{arcsinh}\left(\dfrac{x-\mu}{\sigma}\right)-\varepsilon\right)}{\sigma\sqrt{ 1 + \left(\dfrac{x-\mu}{\sigma}\right)^2 }}$. \cite{JP09} show that density (\ref{SASPDF}) is unimodal and that $(\varepsilon,\delta)$ can be interpreted as skewness and kurtosis parameters, respectively, if they are studied separately. The density (\ref{SASPDF}) contains the normal distribution as a particular case when $(\varepsilon,\delta)=(0,1)$. By fixing $\varepsilon=0$, a symmetric density is obtained with the property that values of $\delta<1$ produce distributions with heavier tails than those of the normal one; values of $\delta>1$ produce distributions with lighter tails. On the other hand, fixing $\delta=1$ yields an asymmetric distribution that contains the normal distribution when $\varepsilon=0$. Another appealing feature is that moments of any order exist for this distribution, for any combination of the parameters. Simulation from this model is straightforward by using the expression (\ref{SASCDF}) together with the probability integral transform. \cite{RJP10} proposed using the sinh--arcsinh transformation $H(x;\mu,\sigma,1,\varepsilon)$ as a method to induce skewness in the Student--$t$ distribution with unknown degrees of freedom. We call this the T SAS distribution. More recently, \cite{FH13} proposed applying the sinh--arcsinh transformation to the hyperbolic secant distribution, in a similar fashion to (\ref{SASPDF}), to produce a flexible distribution centred at the hyperbolic secant distribution. Similarly, \cite{PA14} combined the sinh--arcsinh transformation with the logistic distribution, producing a distribution that can be multimodal.

To quantify the asymmetry levels captured by the SAS distribution, we consider two measures of skewness: (i) the AG measure of skewness \citep{AG95}, which is defined as the difference of the mass cumulated to the right of the mode minus the mass cumulated to the left of the mode, hence taking values in (-1,1); and (ii) the Critchley-Jones (CJ) functional asymmetry measure \citep{CJ08}, which measures discrepancies between points located on either side of the mode $(x_L(p),x_R(p))$ of a density $s$ such that $s(x_L(p))=s(x_R(p))=ps(\mbox{mode})$, $p \in (0,1)$, with formula:

\begin{eqnarray}\label{AsymFun}
\operatorname{CJ}(p)=\dfrac{x_R(p)-2\times\mbox{mode}+x_L(p)}{x_R(p)-x_L(p)}.
\end{eqnarray}

This measure also takes values in $(-1,1)$; negative values of $\operatorname{CJ}(p)$ indicate that the values $x_L(p)$ are further from the mode than the values $x_R(p)$, and analogously for positive values. \cite{CJ08} show that the scalar $\operatorname{AG}$ measure of skewness can be seen as an average of the asymmetry function $\operatorname{CJ}$.

\begin{figure}[h]
\begin{center}
\begin{tabular}{c}
\psfig{figure=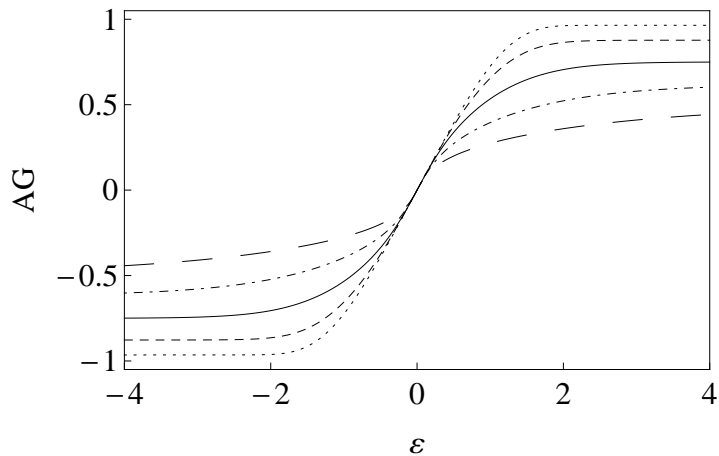,  height=6cm}
\end{tabular}
\end{center}
\caption{\small AG measure of skewness as a function of $\varepsilon$: $\delta=0.25$ (dotted line); $\delta=0.5$ (short dashed line); $\delta=1$ (continuous line); $\delta=2$ (dot-dashed line); $\delta=4$ (long dashed line).}
\label{fig:AGSAS}
\end{figure}

\begin{figure}[h]
\begin{center}
\begin{tabular}{c c}
\psfig{figure=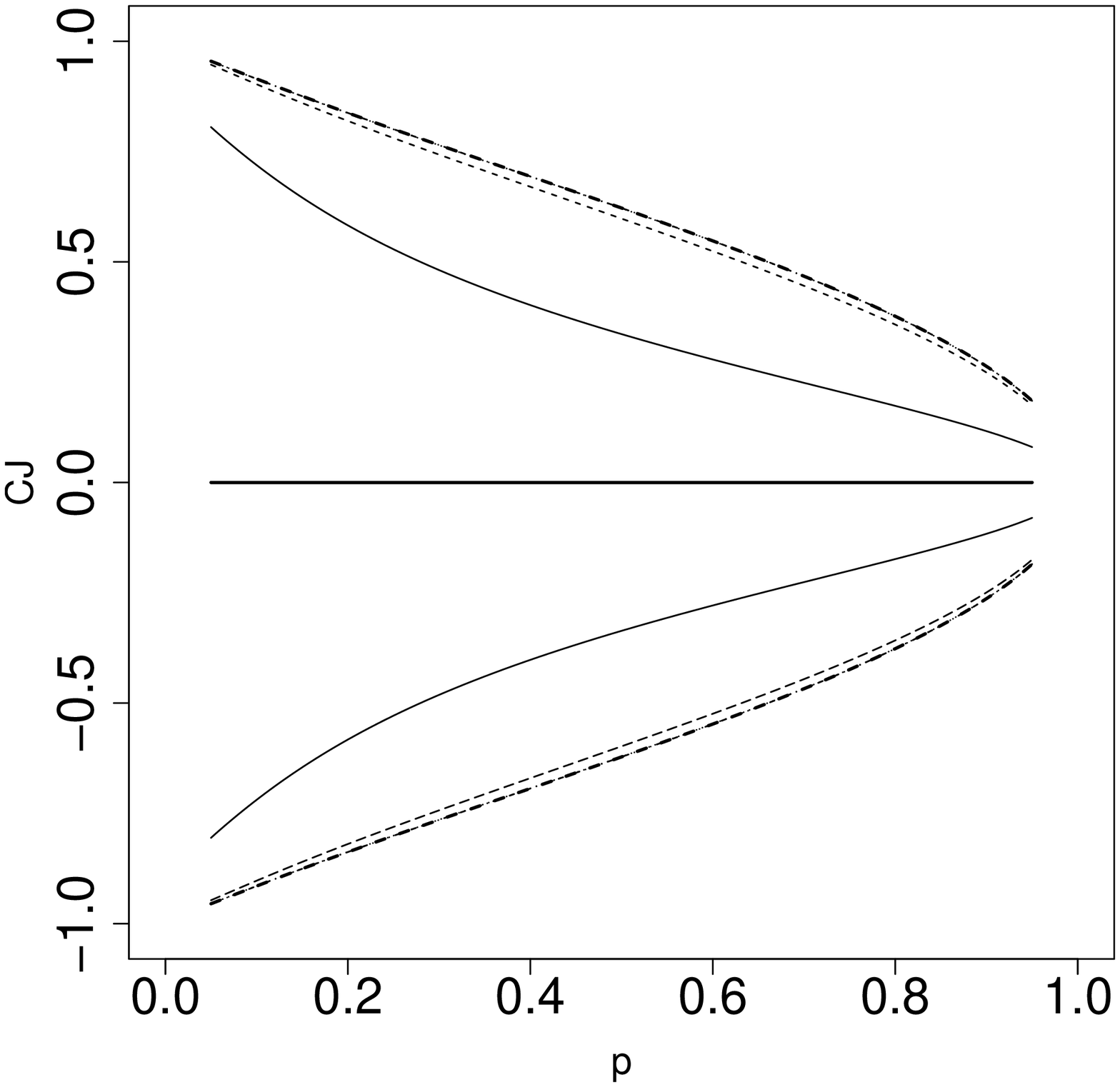,  height=5cm} &
\psfig{figure=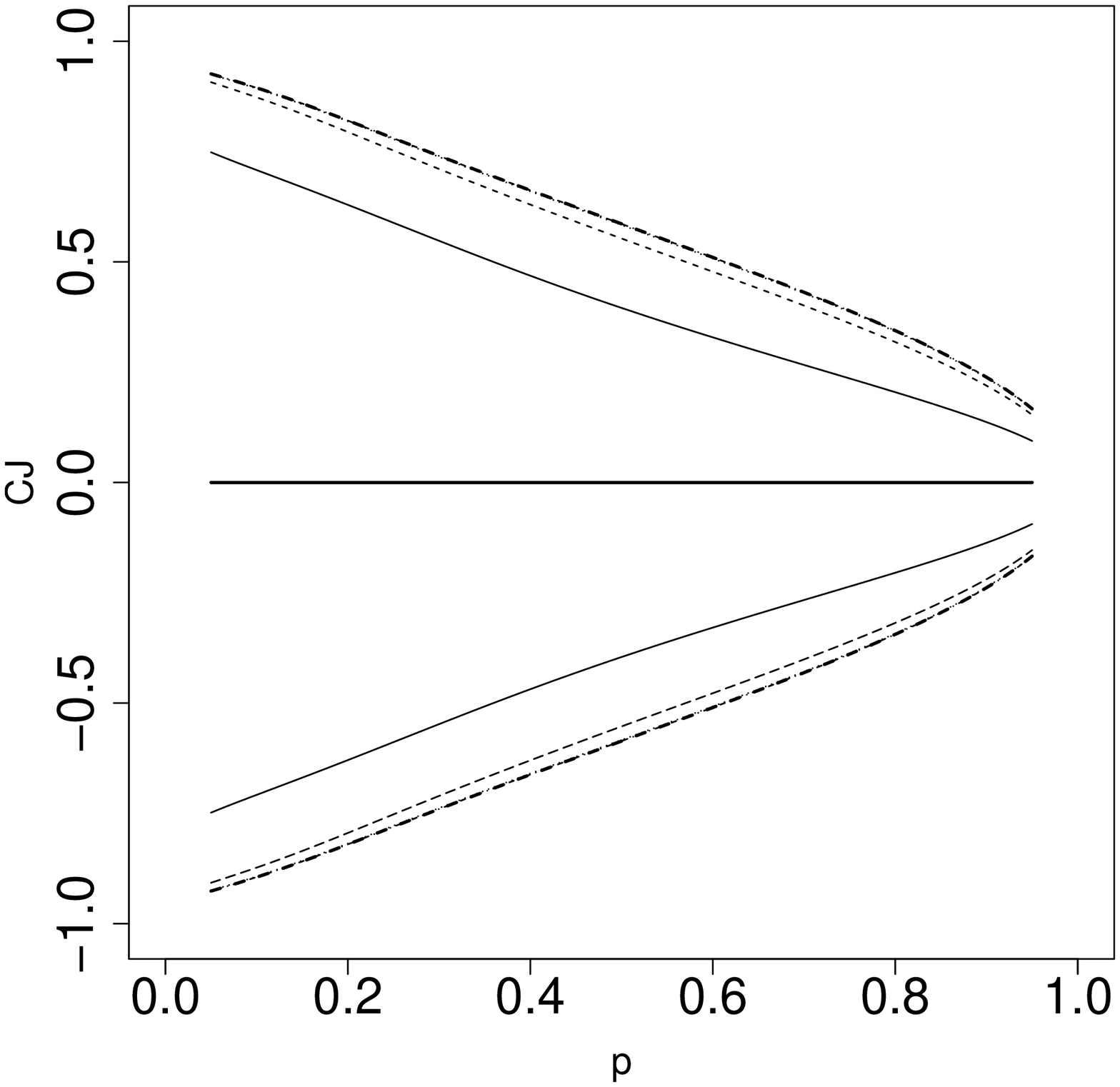,  height=5cm} \\
(a) & (b)\\
\psfig{figure=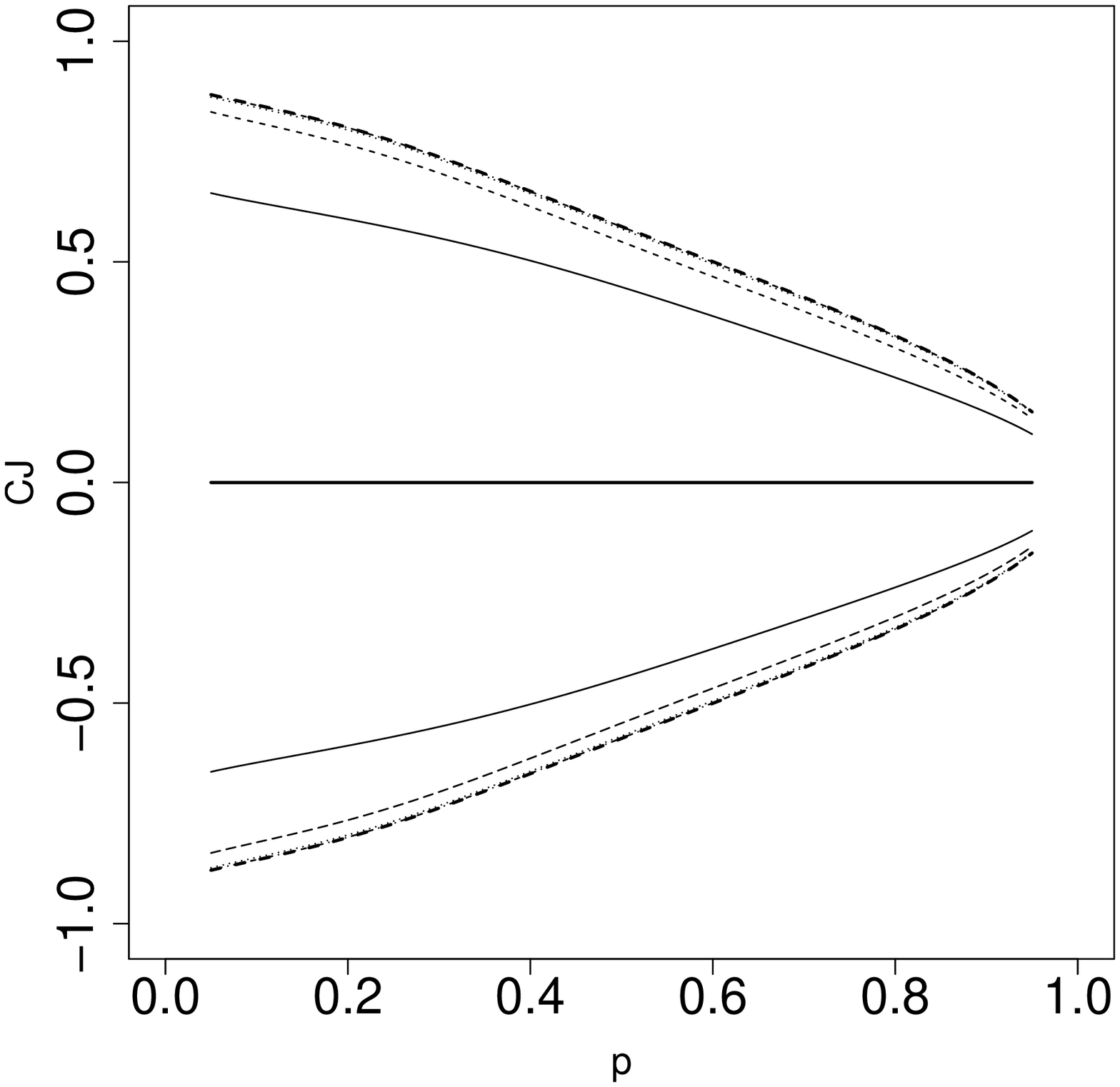,  height=5cm} &
\psfig{figure=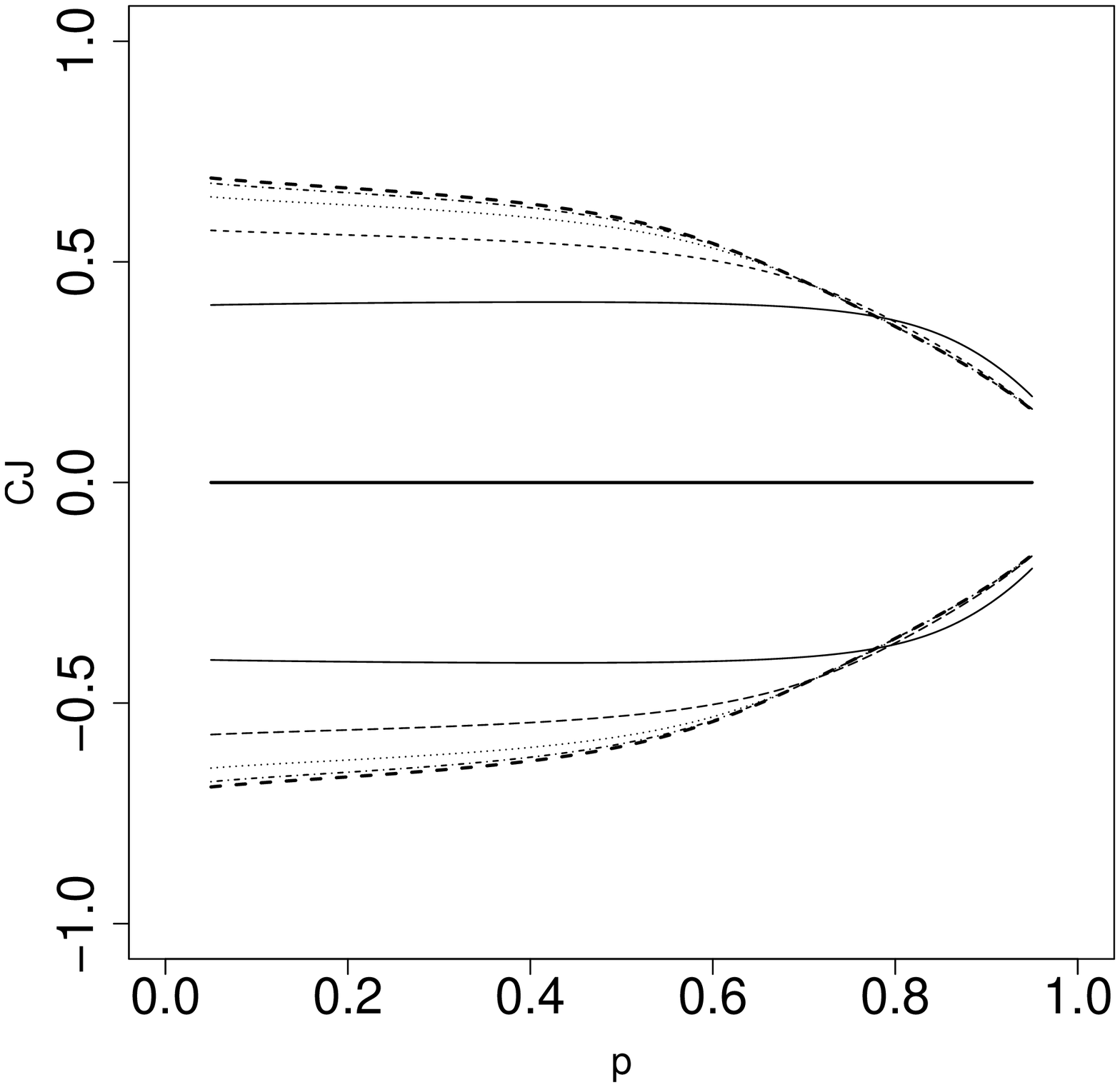,  height=5cm} \\
(c) & (d)\\
\psfig{figure=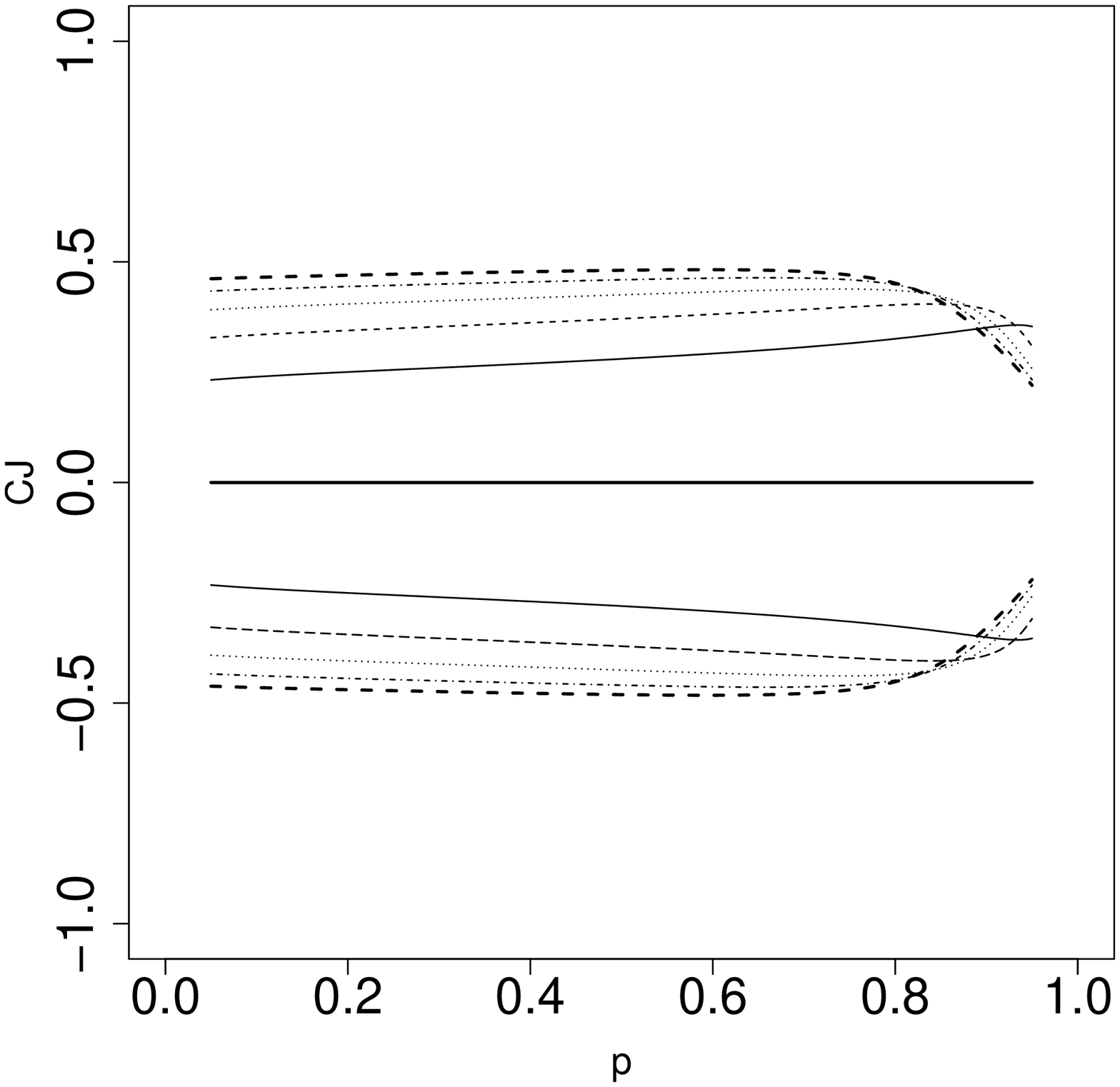,  height=5cm} &
 \\
(e) & \\
\end{tabular}
\end{center}
\caption{\small CJ functional measure of asymmetry: (a) $\delta=0.5$, (b) $\delta=0.75$, (c) $\delta=1$, (d) $\delta=2$, and (e) $\delta=4$. The curves represent the CJ for different values of $\varepsilon$: $\varepsilon=-5,5$ (dashed bold line), $\varepsilon=-4,4$ (dashed-dotted line), $\varepsilon=-3,3$ (dotted line), $\varepsilon=-2,2$ (dashed line), $\varepsilon=-1,1$ (continuous line), and $\varepsilon=0$ (bold line)}
\label{fig:CJSAS}
\end{figure}

Figure \ref{fig:AGSAS} shows the AG measure of (\ref{SASPDF}) obtained by varying the parameter $\varepsilon$ for different values of the parameter $\delta$. This figure indicates that this model covers different ranges of AG for different values of $\delta$, and that these ranges are narrower for larger values of $\delta$. Figure \ref{fig:CJSAS} shows the CJ asymmetry functional measure for different values of $\delta$ and $\varepsilon$. The range of values of CJ covered by varying $\varepsilon$ is also narrower for larger values of $\delta$, and that $\delta$ and $\varepsilon$ have a joint role in controlling the shape of the density.

%

\section{Two--piece sinh--arcsinh distribution}\label{TPSAS}

In order to produce a model that can cover the whole range of the AG and CJ measures of skewness, while keeping some of the original appealing properties of the SAS distribution, we propose a modification obtained by fixing the parameter $\varepsilon=0$ in (\ref{SASPDF}) and then introducing skewness through the \emph{two--piece} transformation.

\begin{definition}
A random variable $X$ is said to be distributed as a two piece sinh--arcsinh (TP SAS) if its pdf is given by:

\begin{eqnarray}\label{twopiecesinhPDF1}
s_1(x;\mu,\sigma_1,\sigma_2,\delta)&=&\dfrac{2}{\sigma_1+\sigma_2}\bigg[f_0\left(\dfrac{x-\mu}{\sigma_1};\delta\right)I(x<\mu) \notag\\
 &+& f_0\left(\dfrac{x-\mu}{\sigma_2};\delta\right)I(x\geq \mu)\bigg],
\end{eqnarray}
\noindent where $f_0(x;\delta)=s_0(x;0,1,0,\delta)$ is the symmetric SAS density, $\mu\in{\mathbb R}$, and $\sigma_1$, $\sigma_2$, $\delta\in{\mathbb R}_+$.
\end{definition}

\begin{figure}[h]
\begin{center}
\begin{tabular}{c c c}
\psfig{figure=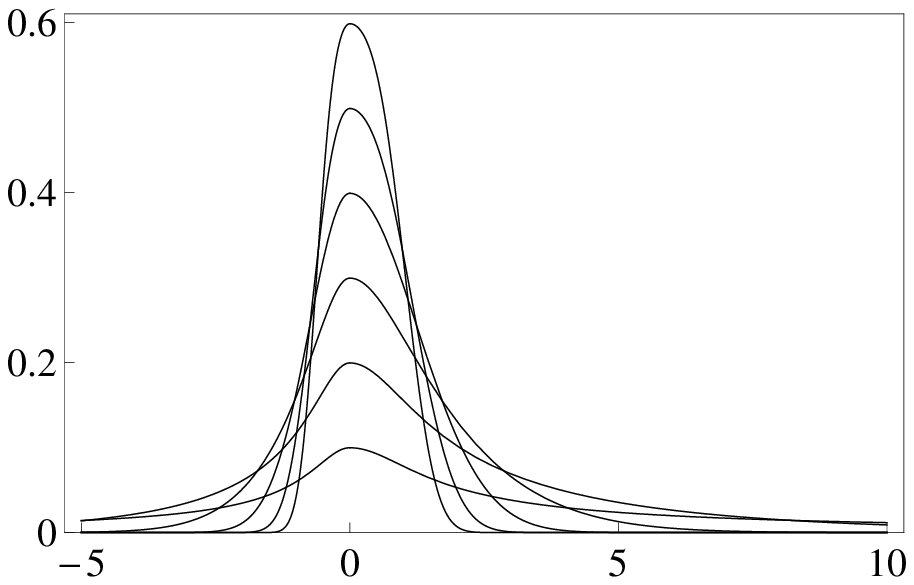,  height=3cm} &
\psfig{figure=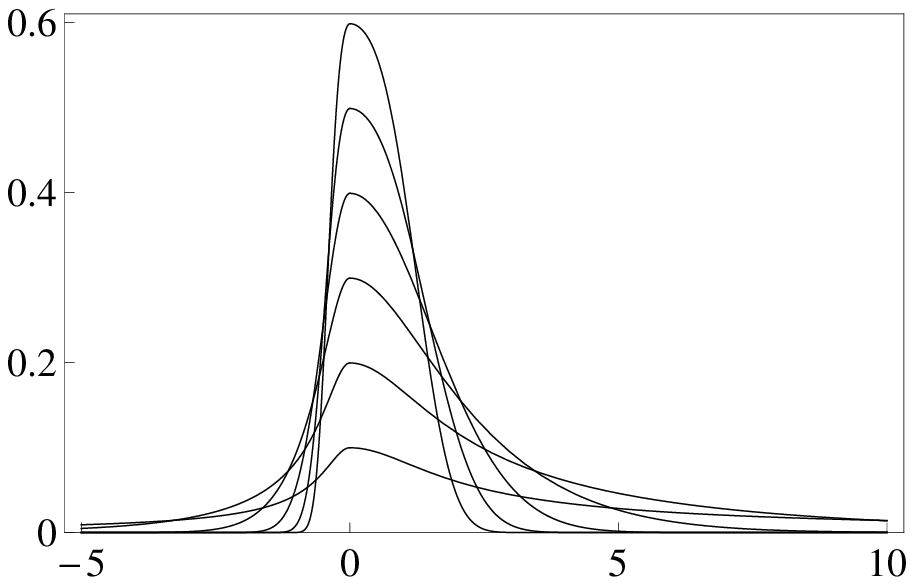,  height=3cm} &
\psfig{figure=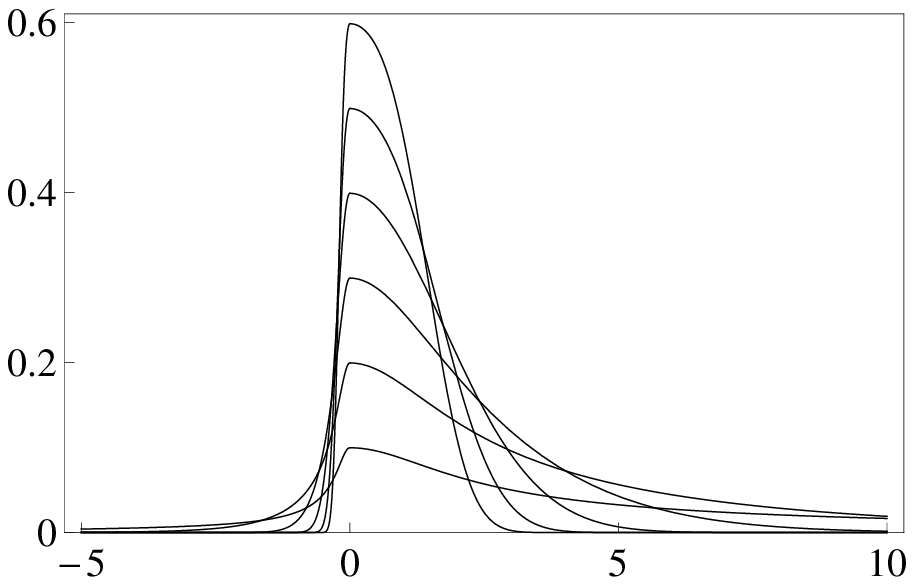,  height=3cm}\\
(a) & (b) & (c)
\end{tabular}
\end{center}
\caption{\small Two--piece sinh--arcsinh density $(\ref{twopiecesinhPDF})$ under the $\epsilon-$skew parameterisation with $\mu=0$, $\sigma=1$, $\delta=(0.25,0.5,\dots,1.5)$ and: (a) $\gamma=-0.25$; (b) $\gamma=-0.5$; (c) $\gamma=-0.75$.}
\label{fig:TPshapes}
\end{figure}

The density (\ref{twopiecesinhPDF1}) joins two symmetric SAS half--densities at the mode with different scale parameters. This pdf is unimodal, with mode at $\mu$, contains the symmetric SAS distribution for $\sigma_1=\sigma_2$, and is asymmetric for $\sigma_1\neq \sigma_2$. Given that the symmetric SAS distribution is an identifiable model \citep{JP09}, it follows that the TP SAS distribution is identifiable as well. Moreover, the tail behaviour of the TP SAS distribution is the same in each direction given that it is obtained as a transformation of scale \citep{J14}. A useful family of reparameterisations of distributions of the type (\ref{twopiecesinhPDF1}) was proposed by \cite{AValle05} as follows:

\begin{eqnarray}\label{twopiecesinhPDF}
s_1(x;\mu,\sigma,\gamma,\delta)&=&\dfrac{2}{\sigma[a(\gamma)+b(\gamma)]}\bigg[f_0\left(\dfrac{x-\mu}{\sigma b(\gamma)};
\delta\right)I(x<\mu) \notag\\
 &+& f_0\left(\dfrac{x-\mu}{\sigma a(\gamma)};\delta\right)I(x\geq \mu)\bigg],
\end{eqnarray}

\noindent where $a(\gamma)>0$, $b(\gamma)>0$, $\gamma \in \Gamma$. The space $\Gamma$ depends on the parameterisation $\{a(\gamma),b(\gamma)\}$. Perhaps, the most popular parameterisations correspond to the cases when $\{a(\gamma),b(\gamma)\}=\{1/\gamma,\gamma\}$, $\gamma>0$, termed \emph{inverse scale--factors} parameterisation \citep{FS98}, and $\{a(\gamma),b(\gamma)\}=\{1-\gamma,1+\gamma\}$, $\gamma\in(-1,1)$, termed \emph{$\epsilon-$skew} parameterisation \citep{MH00}. Some other parameterisations were studied in \cite{RS14a}. For $\delta=1$, we obtain the two--piece normal distribution \citep{MH00}, and for $a(\gamma)=b(\gamma)$ we obtain the symmetric sinh--arcsinh distribution \citep{JP09}. Figure \ref{fig:TPshapes} shows some examples of the shapes of density (\ref{twopiecesinhPDF}) for the \emph{$\epsilon-$skew} parameterisation.

For two-piece distributions, \cite{KF06} showed that the parameter $\gamma$ can be interpreted as a skewness parameter in a more fundamental sense (often called ``van Zwet ordering'', \citealp{Zwet64}). This means that $(\gamma,\delta)$ in (\ref{twopiecesinhPDF}) can also be interpreted as skewness and kurtosis parameters, respectively, in the same way that \cite{JP09} interpreted $(\varepsilon,\delta)$ for the SAS distribution. The AG and the CJ measures coincide for this model, and depend only on $\gamma$, as follows:

\begin{eqnarray*}
\mbox{AG}(\gamma) = \mbox{CJ}(p,\gamma) = \frac{a(\gamma)-b(\gamma)}{a(\gamma)+b(\gamma)}.
\end{eqnarray*}

For instance, under the $\epsilon-$skew parameterisation $\text{AG}(\gamma)= \mbox{CJ}(p,\gamma)=-\gamma \in(-1,1)$. From this expression, it is clear that model (\ref{twopiecesinhPDF}) includes the whole range of AG and CJ measures.

One important difference between models (\ref{SASPDF}) and (\ref{twopiecesinhPDF}) is that in (\ref{SASPDF}), the parameter $\varepsilon$ also controls the tail behaviour. In fact, values of $\varepsilon\neq 0$ produce asymmetric densities with different tail behaviour in each direction \citep{JP09}. This type of asymmetry (with different tails) was recently denoted ``tail asymmetry'' by \cite{J14}. On the other hand, (\ref{twopiecesinhPDF}) has the same tail behaviour in each direction, denoted ``main-body asymmetry'' by \cite{J14}. This difference between the TP SAS and SAS distributions is neither an advantage of one over the other nor a disadvantage: these models capture different types of asymmetry. However, in practice, the data may favour one of these types of asymmetry. Therefore, a model comparison between the TP SAS and SAS models could also provide information about the type of asymmetry that better fits the data. Distributions that can capture both types of asymmetry have been recently studied in \cite{RS14b}.

\subsection{Some properties of the two-piece sinh--arcsinh distribution}

We now discuss some basic properties of the TP SAS distribution which show the tractability of this model. These properties are largely inherited from the well-known properties of the two-piece transformation.

The cdf of the TP SAS distribution is given by the following expression.


\begin{eqnarray}\label{twopiecesinhCDF}
S_1(x;\mu,\sigma,\gamma,\delta) &=& \dfrac{2b(\gamma)}{a(\gamma)+b(\gamma)} F_0\left(\dfrac{x-\mu}{\sigma b(\gamma)};\delta\right) I(x<\mu) \\
&+& \left[\frac{b(\gamma)-a(\gamma)}{a(\gamma)+b(\gamma)} + \frac{2a(\gamma)}{a(\gamma) + b(\gamma)}F_0\left(\dfrac{x-\mu}{\sigma a(\gamma)};\delta\right)\right]I(x\geq \mu),\notag
\end{eqnarray}

\noindent where $F_0(x;\delta)=S_0(x;0,1,0,\delta)$ is the symmetric SAS cdf. The quantile function can be easily obtained by inverting this expression.

{\bf Moments.} Given that the moments of any order of the symmetric SAS distribution exist \citep{JP09}, and that the two--piece transformation preserves the existence of moments \citep{AValle05}, it follows that moments of any order of the TP SAS distribution (\ref{twopiecesinhCDF}) exist, for any combination of the parameters. Expressions for the moments of (\ref{twopiecesinhPDF}) can be derived by combining the expressions for the moments of the symmetric SAS distribution from \cite{JP09} and the expression for the moments of two-piece distributions in \cite{AValle05}. However, these expressions are slightly cumbersome and difficult to interpret. The moments can be fairly easily calculated using numerical integration, so we do not give the formulae.

{\bf Inference.} Although two--piece distributions are not twice differentiable at the mode, a (sufficient) regularity condition required in some classical results, this feature does not preclude ML estimation methods in this family. Asymptotic results (consistency and asymptotic normality) for ML estimators have been obtained using direct proofs (in some specific cases) \citep{MH00,AValle05}. \cite{JA10} also show that certain parameterisations of the two--piece family of distributions, such as the $\epsilon-$skew parameterisation, induce partial parameter orthogonality which improves some asymptotic properties of ML estimators. The expression for ML estimators of the TP SAS distribution is not available in closed-form, hence numerical methods are required.

{\bf Multivariate Extensions.} Although there is no ``natural'' extension of the two-piece transformation to the multivariate case, multivariate extensions of these models have been explored using Copulas \citep{RS13} and affine transformations \citep{FS07}. \cite{BL05} propose a method to construct $2^k$-piece distributions which can be applied to $k-$variate distributions with a certain type of symmetry. These ideas can be immediately applied to the TP SAS distribution in order to produce multivariate extensions of the model.

\subsection{Skew--symmetric sinh--arcsinh distribution}

Since our motivation for introducing the TP SAS distribution consists of producing a model that can cover the whole range of some interpretable measures of asymmetry for any value of $\delta$, an immediate question is whether  there are other transformations for doing so. The answer is positive, the skew-symmetric construction being a natural candidate among the most popular skewing mechanisms. Other popular density-based transformations such as the Marshall-Olkin and the power transformations have been shown to induce little flexibility on the symmetric SAS distribution (Chapter 2, \citealp{R13}).

\begin{definition}
A random variable is said to be distributed as a skew--symmetric sinh--arcsinh (SS SAS) if its pdf is given by:

\begin{eqnarray}\label{SSsinhPDF}
s_2(x;\mu,\sigma,\lambda,\delta)&=& \dfrac{2}{\sigma} f_0\left(\dfrac{x-\mu}{\sigma};\delta\right) F_0\left(\lambda\dfrac{x-\mu}{\sigma };\delta\right),
\end{eqnarray}
\noindent where $\mu,\lambda \in{\mathbb R}$, and $\sigma,\delta\in{\mathbb R}_+$.
\end{definition}

\begin{figure}[h]
\begin{center}
\begin{tabular}{c c c}
\psfig{figure=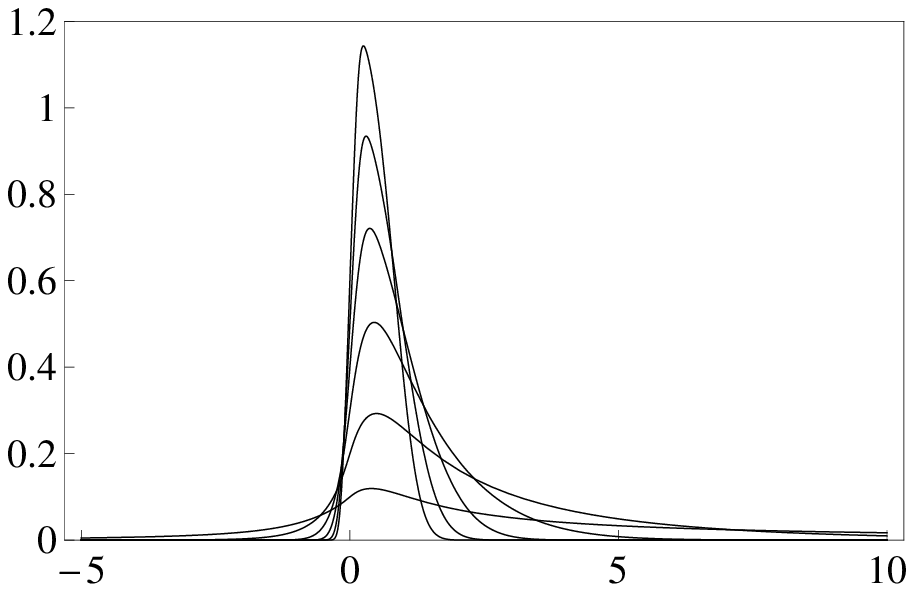,  height=3cm} &
\psfig{figure=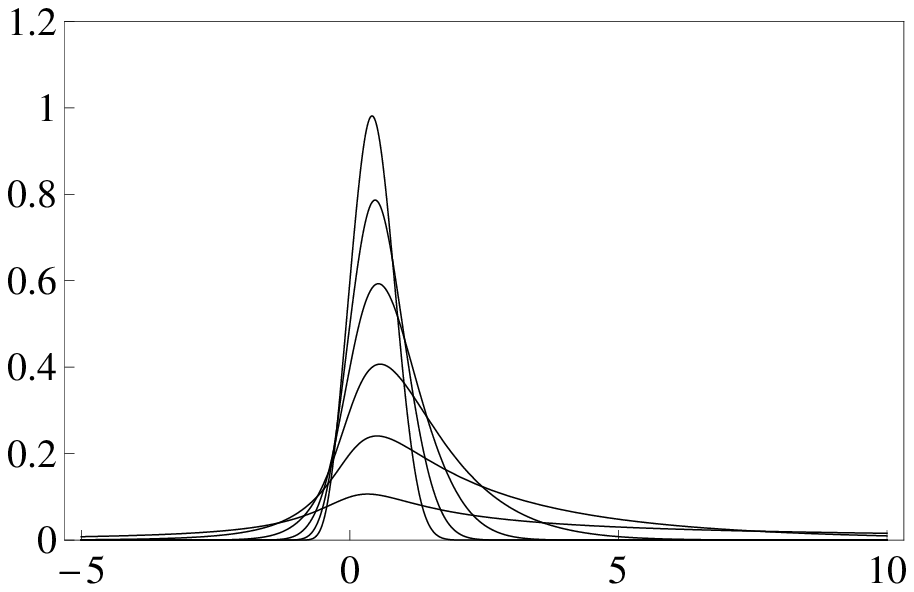,  height=3cm} &
\psfig{figure=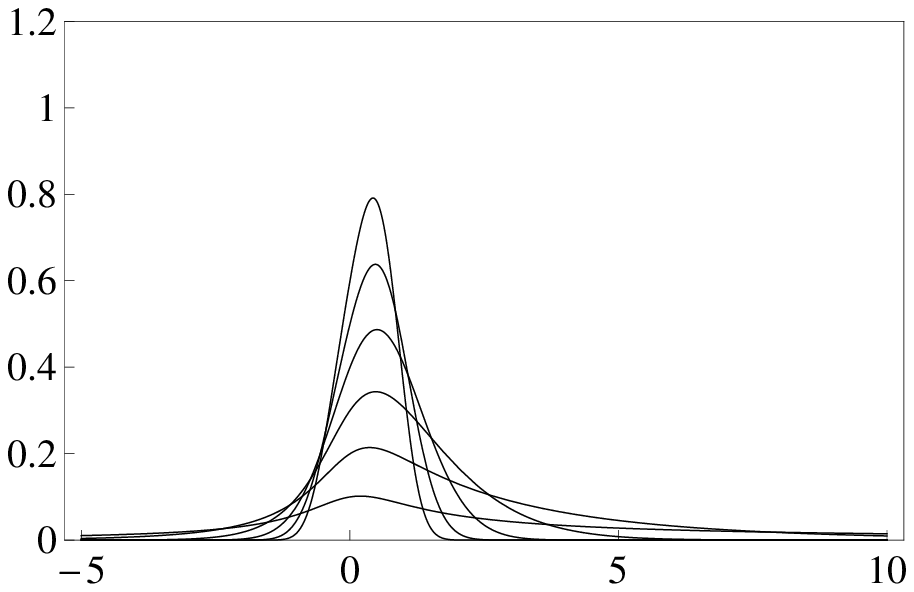,  height=3cm}\\
(a) & (b) & (c)
\end{tabular}
\end{center}
\caption{\small Skew--symmetric sinh--arcsinh density $(\ref{SSsinhPDF})$ with $\mu=0$, $\sigma=1$, $\delta=(0.25,0.5,\dots,1.5)$ and: (a) $\lambda=1$; (b) $\lambda=2$; (c) $\lambda=5$.}
\label{fig:SSshapes}
\end{figure}

\begin{figure}[h]
\begin{center}
\begin{tabular}{c c}
\psfig{figure=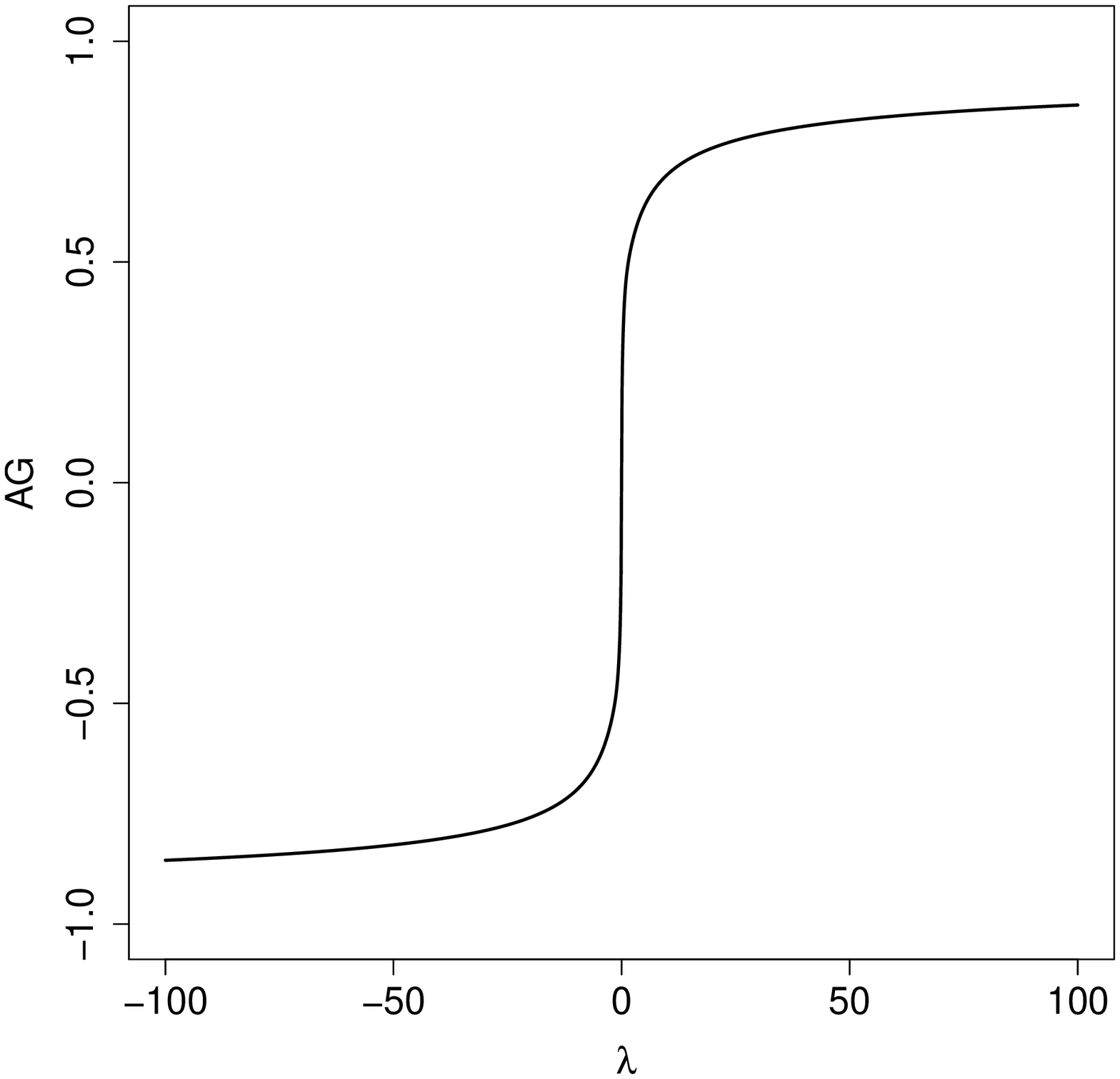,  height=5cm} &
\psfig{figure=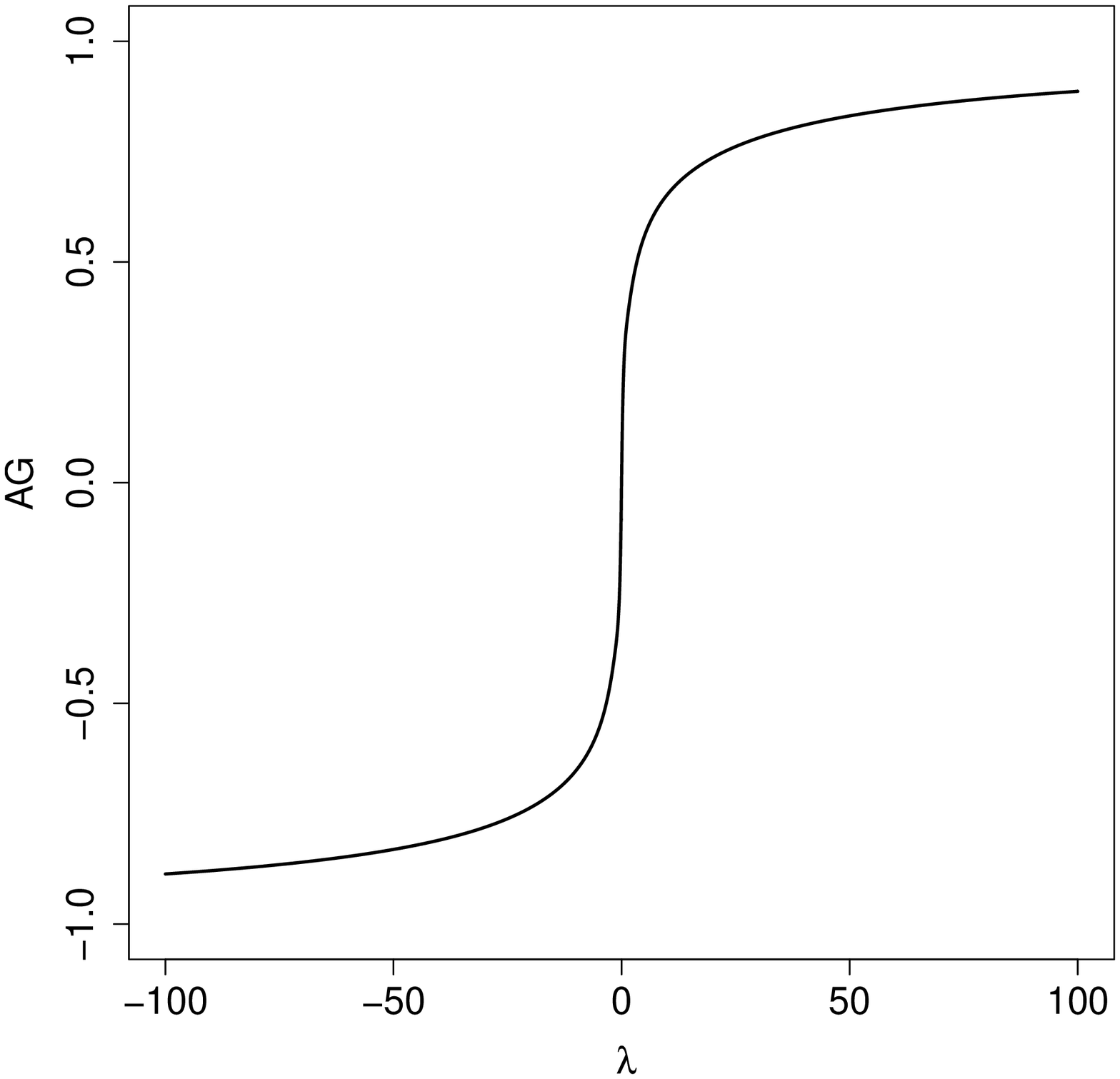,  height=5cm} \\
(a) & (b)\\
\psfig{figure=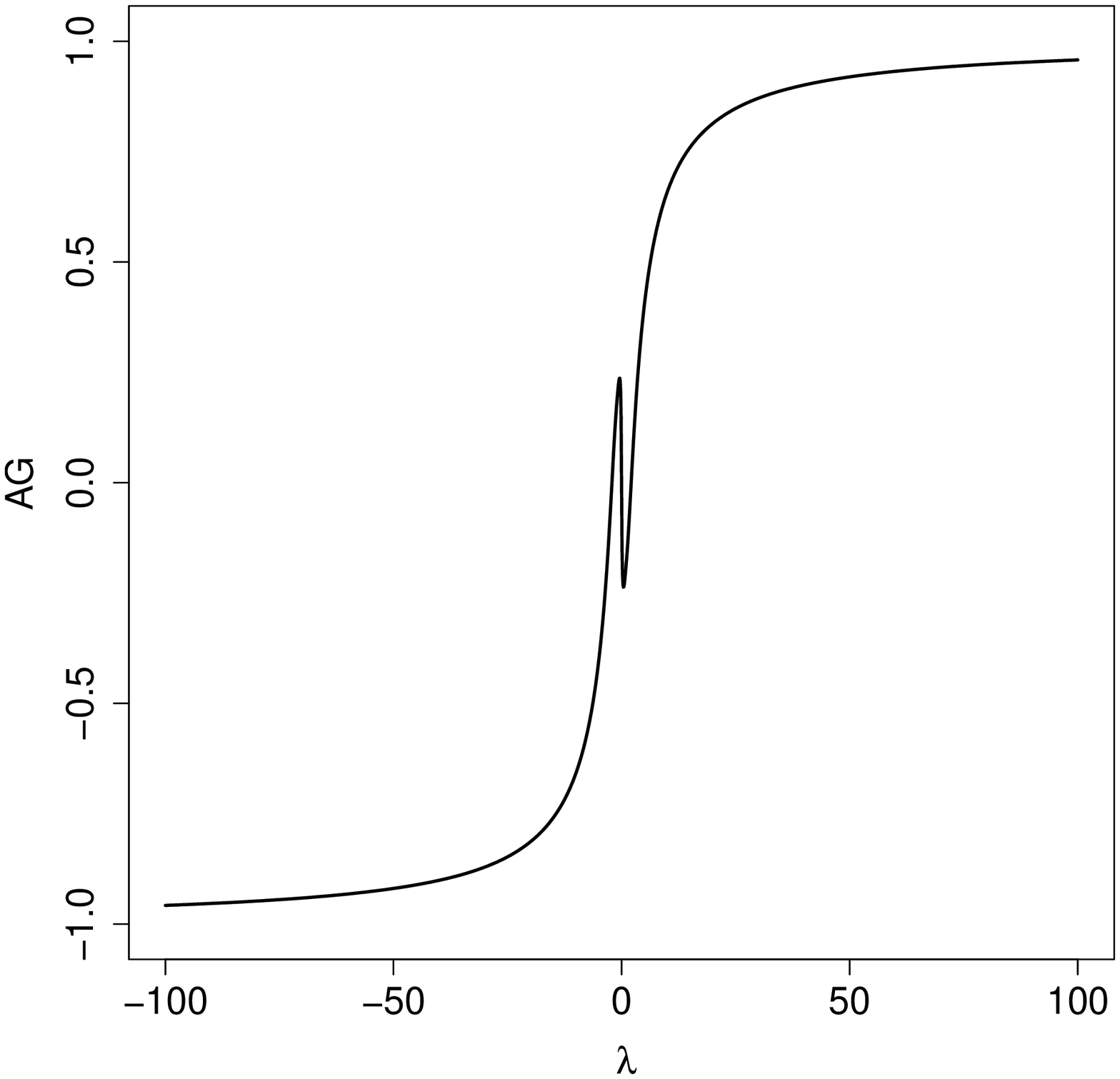,  height=5cm} &
\psfig{figure=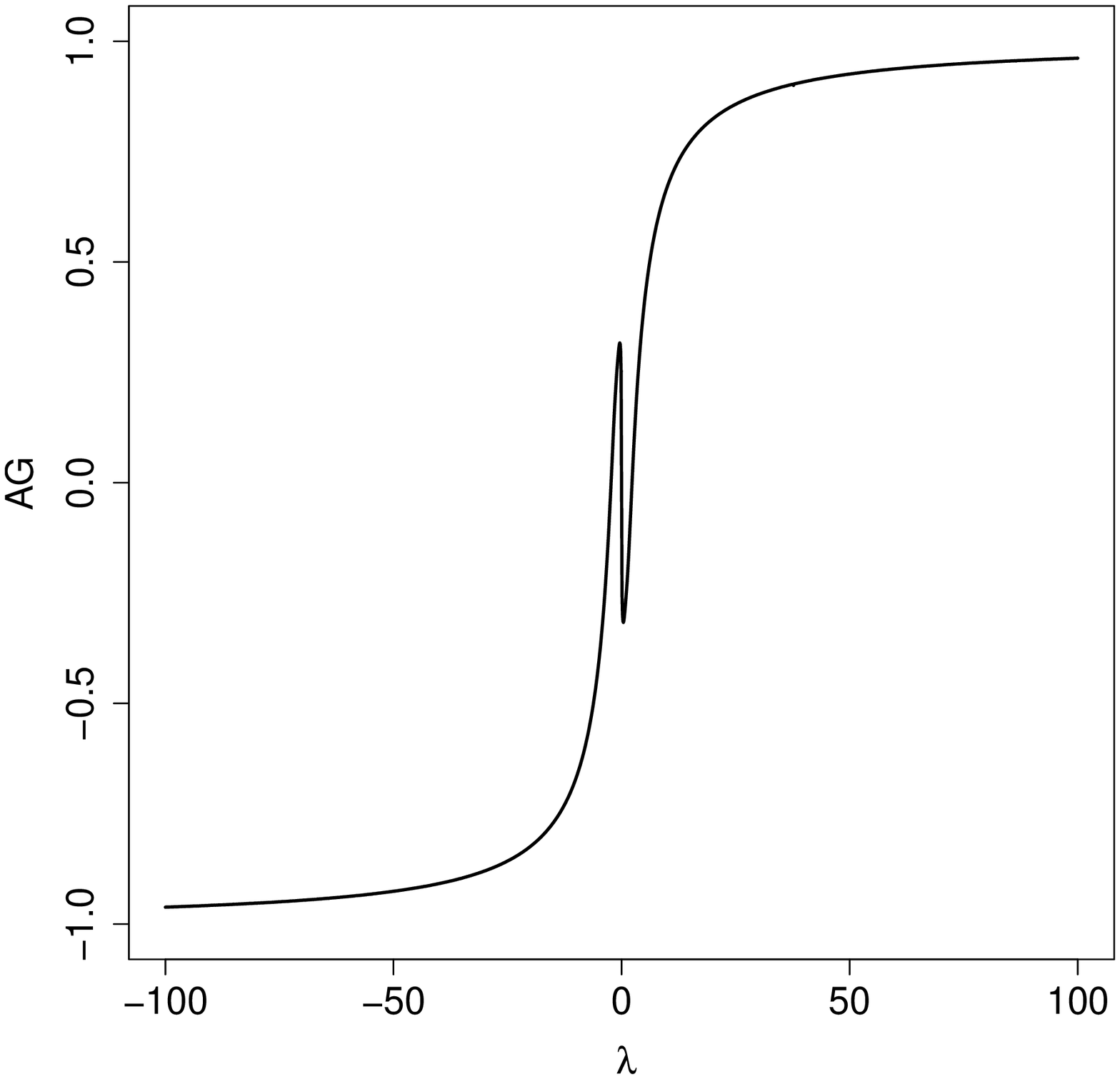,  height=5cm} \\
(c) & (d)
\end{tabular}
\end{center}
\caption{\small AG measure of skewness as a function of $\lambda$: (a) $\delta=0.25$; (b) $\delta=0.5$; (c) $\delta=1$; (d) $\delta=4$.}
\label{fig:AGSS}
\end{figure}

This density contains the symmetric SAS distribution for $\lambda=0$, it is asymmetric for $\lambda\neq 0$, and it converges to the right/left half symmetric SAS as $\lambda \rightarrow \pm \infty$. This property is typically used to interpret $\lambda$ as a skewness parameter, and it also implies that the model can cover the full range of the AG measure. However, AG is not an injective function of the parameter $\lambda$  for $\delta\geq 1$, as shown in Figure \ref{fig:AGSS}. Figure \ref{fig:SSshapes} shows the shapes of this model for different values of the parameters. We can observe that the parameter $\lambda$ also controls the mode and the tails of the density. In fact, it can be shown that the distribution has different tails in each direction, a property shared by all skew-symmetric distributions, implying also that the SS SAS capture ``tail asymmetry''. Even though the SS SAS covers the whole range of AG, it also inherits all the inferential properties of the skew-normal distribution \citep{A85}, which is a particular case of (\ref{SSsinhPDF}). This might represent a drawback for some practitioners given the inferential problems with the skew-normal discussed earlier. However, these problems mainly related to small samples (see \citealp{J14} for a discussion on this point).

\section{A short simulation study}

We conducted a short simulation study to evaluate the performance of the ML estimators of the parameters of the TP SAS distribution. We simulated $N=10,000$ samples from a TP SAS distribution (with the $\epsilon-$skew parameterisation) for a range of parameter values and sample sizes, and calculated the bias, variance and root-mean-square error (RMSE) of the ML estimators for each scenario. Results are presented in Tables \ref{table:SIMTPSAS1} and \ref{table:SIMTPSAS2}. The simulation study reveals that the skewness level does not seem to affect the behaviour of the ML estimators. However, the tail parameter is clearly difficult to estimate whether the samples come from a distribution with lighter or heavier tails than normal. The results suggest that lighter tails are harder to estimate in the sense that larger samples are required to accurately estimate the tail parameter. This is an intriguing behaviour that require further general research. We would like to quote a discussion from \cite{J14} with respect to this point: ``I suspect we do not understand very light-tailed distributions very well, perhaps reasonably so given their relative scarcity in practice''. The simulation suggested that to estimate the tail parameter accurately we may need at least a couple of hundred observations. This is not a surprising phenomenon since tail parameters are known to be difficult to estimate, such as the tail parameters in the Student-$t$ distribution, the exponential power distribution and generalised hyperbolic distribution \citep{FMF12}.

\clearpage
\afterpage{\clearpage
\begin{landscape}
\vspace*{\fill}
\begin{table}[h!]
\begin{center}
\begin{tabular}[h]{|c|c|c|c|c|c|c|c|c|c|c|c|c|c|c|c|c|c|}
\hline
\multicolumn{2}{|c|}{Parameters} & $\mu=0$ & $\sigma=1$ & $\gamma=0.25$ & $\delta=1.25$ & $\mu=0$ & $\sigma=1$ & $\gamma=0.5$ & $\delta=1.25$ & $\mu=0$ & $\sigma=1$ & $\gamma=0.75$ & $\delta=1.25$\\
\hline
 & $n$ & $\hat\mu$ & $\hat\sigma$ & $\hat\gamma$ & $\hat\delta$ & $\hat\mu$ & $\hat\sigma$ & $\hat\gamma$ & $\hat\delta$ & $\hat\mu$ & $\hat\sigma$ & $\hat\gamma$ & $\hat\delta$ \\
\hline
\multirow{5}{*}{Bias} &  50 & -0.039 & -962.5 & -0.040 & -1075  &   -0.068  &  -531.7  &  -0.073  & -586.5  &  -0.052  & -573.6 & -0.075  &  -638.8 \\
& 100  & -0.027 & -287.7 &  -0.023 & -310.5  & -0.046   &  -299.9  &  -0.042  & -327.5  &  -0.047  & -112.6 & -0.050  &  -129.92 \\
& 250 & -0.005 & -5.2 &  -0.005  &  -5.566  & -0.011  & -2.24 &  -0.010  & -2.41  &  -0.016  & -1.33 & -0.016  & -1.45  \\
& 500 & -0.001 & -0.049 & -0.001 & -0.051  &  -0.004  & -0.050   & -0.004   &  -0.052 &  -0.006  & -0.051 & -0.006  & -0.054  \\
& 1000 & -2.7$\times10^{-5}$ & -0.002 & -2.5$\times10^{-4}$ & -0.021  & -9.0$\times10^{-4}$   &  -0.020  &  -0.001  & -0.021  &  -0.002  & -0.020 &  -0.002 &  -0.021 \\
\hline
\multirow{5}{*}{Var.} &  50 & 0.234 & 4.2$\times 10^8$ & 0.149 & 5.4 $\times10^8$  &  0.157  & 7.9$\times10^7$   &  0.101  & 1.1$\times10^8$  &  0.007  & 2.5$\times10^8$ & 0.048  &  3.1$\times10^8$ \\
& 100  & 0.083 & 5.3$\times10^7$ & 0.049 & 6.1$\times10^7$  & 0.067   &  1.0$\times10^8$  & 0.040   & 1.2$\times10^8$  & 0.037   & 2.3$\times10^7$ &   0.023& 3.2$\times10^7$  \\
& 250 & 0.022 & 6.4$\times10^4$ & 0.001 & 7.1$\times10^4$  &  0.002  &  8.8$\times10^3$  &   0.001 & 1.1$\times10^4$  &  0.013  & 3.1$\times10^3$ &  0.008 & 3.6$\times10^3$  \\
& 500 & 0.010 & 0.043 & 0.005 &  0.039 & 0.008   &  0.044   &  0.004  &  0.041 & 0.005   & 0.046 &  0.003 & 0.043  \\
& 1000 & 0.005 & 0.016 & 0.002 & 0.014  &  0.004  &  0.016  &  0.002  &  0.015 &  0.002  & 0.016 & 0.001  & 0.014  \\
\hline
\multirow{5}{*}{RMSE} &  50 & 0.486 & 2.0$\times10^4$ & 0.388 & 2.3$\times10^4$  & 0.402   &  8.9$\times10^3$  & 0.327   & 1$\times10^4$  &  0.277  & 1.6$\times10^4$ &  0.232 & 1.8$\times10^4$  \\
& 100  & 0.289 & 7323.8 & 0.222 &  7824.8 & 0.264   &  1.0$\times10^4$  &  0.205  & 1.1$\times10^4$  &  0.198  & 4825 & 0.159  & 5679  \\
& 250 & 0.148 &  252.9 & 0.111 &  265.2 &  0.139  &  94.08  &  0.104  &  103.0 &   0.117 &  55.3 & 0.089  &  60.4 \\
& 500 & 0.101 & 0.213 &0.075  &  0.205 &  0.092  &  0.217  & 0.069   &  0.209 &  0.073  & 0.222 &  0.055 &   0.214 \\
& 1000 & 0.071 & 0.129 & 0.052 &  0.121 &  0.064  &  0.130  &  0.047  & 0.123  & 0.050   & 0.130 & 0.037  & 0.123  \\
\hline
\end{tabular}
\caption{\small Behaviour of ML estimators for the TP SAS distribution; $\delta>1$, light tails.}
\label{table:SIMTPSAS1}
\end{center}
\end{table}
\vspace*{\fill}
\end{landscape}
}

\afterpage{\clearpage
\begin{landscape}
\vspace*{\fill}
\begin{table}[h!]
\begin{center}
\begin{tabular}[h]{|c|c|c|c|c|c|c|c|c|c|c|c|c|c|c|c|c|c|}
\hline
\multicolumn{2}{|c|}{Parameters} & $\mu=0$ & $\sigma=1$ & $\gamma=0.25$ & $\delta=0.75$ & $\mu=0$ & $\sigma=1$ & $\gamma=0.5$ & $\delta=0.75$ & $\mu=0$ & $\sigma=1$ & $\gamma=0.75$ & $\delta=0.75$\\
\hline
 & $n$ & $\hat\mu$ & $\hat\sigma$ & $\hat\gamma$ & $\hat\delta$ & $\hat\mu$ & $\hat\sigma$ & $\hat\gamma$ & $\hat\delta$ & $\hat\mu$ & $\hat\sigma$ & $\hat\gamma$ & $\hat\delta$ \\
\hline
\multirow{5}{*}{Bias} &  50 & -0.043 & -91.6 & -0.023 & -51.69  &  -0.071  & 71.1   &  -0.042  & -37.9  & -0.058   & -11.33 & -0.050  & -5.82  \\
& 100  & -0.018 & -0.415 &  -0.009& -0.219  &  -0.033  &  -0.509  &  -0.018  & -0.259  & -0.037   & -0.232 & -0.025  & -0.126  \\
& 250 & -0.003 & -0.035 & -0.002 & -0.018  &  -0.006  &  -0.035  &  -0.005  &  -0.018 &  -0.009  & -0.034 & -0.007  & -0.018  \\
& 500 & -0.001 &  -0.016 &  -0.001 & -0.008  &   -0.003 &  -0.016  & -0.002   & -0.008 &   -0.003  & -0.017 &  -0.003 &  -0.008   \\
& 1000 & 4$\times10^{-4}$ & -0.006 & -2.8$\times10^{-5}$ & 0.004  &  7$\times10^{-5}$  & -0.006   & 5$\times10^{-4}$   & -0.003  &  -7$\times10^4$  & -0.007 & -0.001  & -0.003 \\
\hline
\multirow{5}{*}{Var.} &  50 & 0.384 & 1.7$\times10^7$ & 0.068 & 6.4$\times10^6$  &  0.266  & 9.5$\times10^6$   & 0.049   & 2.6$\times10^6$  &  0.129  & 2.6$\times10^5$ & 0.025  &  5.8$\times10^4$ \\
& 100  &  0.114 & 232.8 & 0.020 & 69.93  & 0.098   &   856.7  &   0.017  &  215.0 &  0.063  & 96.47 &  0.011 & 29.90  \\
& 250 & 0.034 & 0.039 & 0.006 &  0.007 & 0.029   &   0.039  &  0.005  & 0.007  &  0.019  & 0.039 & 0.003  &  0.007 \\
& 500 & 0.017 & 0.016 &  0.003 & 0.003  &  0.014  &  0.016  &  0.002  & 0.003  & 0.008   & 0.016 &  0.001 &   0.003 \\
& 1000 & 0.008 & 0.007 & 0.001 & 0.001  &  0.006  &   0.007 &  0.001  &  0.001 & 0.004   & 0.007 & 7$\times10^4$  & 0.001  \\
\hline
\multirow{5}{*}{RMSE} &  50 & 0.621 & 4139 &  0.262 & 2533  & 0.521   &  3085  &  0.227  & 1618  &  0.364  & 513.8 & 0.167  & 241.8 \\
& 100  & 0.338 & 15.2 & 0.144 & 8.36  &  0.315  &  29.27  & 0.135   & 14.66  &  0.254  & 9.82 &  0.110 &  5.47 \\
& 250 & 0.186 & 0.201 & 0.080 & 0.087  &  0.171  & 0.200   &  0.073  &  0.086 &  0.140  & 0.201 & 0.059  & 0.087  \\
& 500 & 0.130 & 0.130 & 0.055 &  0.055 &  0.118  &  0.130  &  0.050  & 0.055  &  0.093  & 0.130 & 0.039  & 0.055  \\
& 1000 & 0.091 & 0.087 & 0.038 &  0.037 &  0.082  &  0.087  &  0.035  & 0.037  &  0.064  & 0.088 & 0.027  & 0.037  \\
\hline
\end{tabular}
\caption{\small Behaviour of ML estimators for the TP SAS distribution; $\delta<1$, heavy tails.}
\label{table:SIMTPSAS2}
\end{center}
\end{table}
\vspace*{\fill}
\end{landscape}
}

\clearpage
\section{Illustrative Example: Internet Traffic Data}\label{Examples}

Data on internet traffic is analysed to illustrate and compare the performance of the SAS, TP SAS, and SS SAS distributions. For the TP SAS model we adopt the $\epsilon-$skew parameterisation.

\begin{figure}[h]
\begin{center}
\begin{tabular}{c c}
\psfig{figure=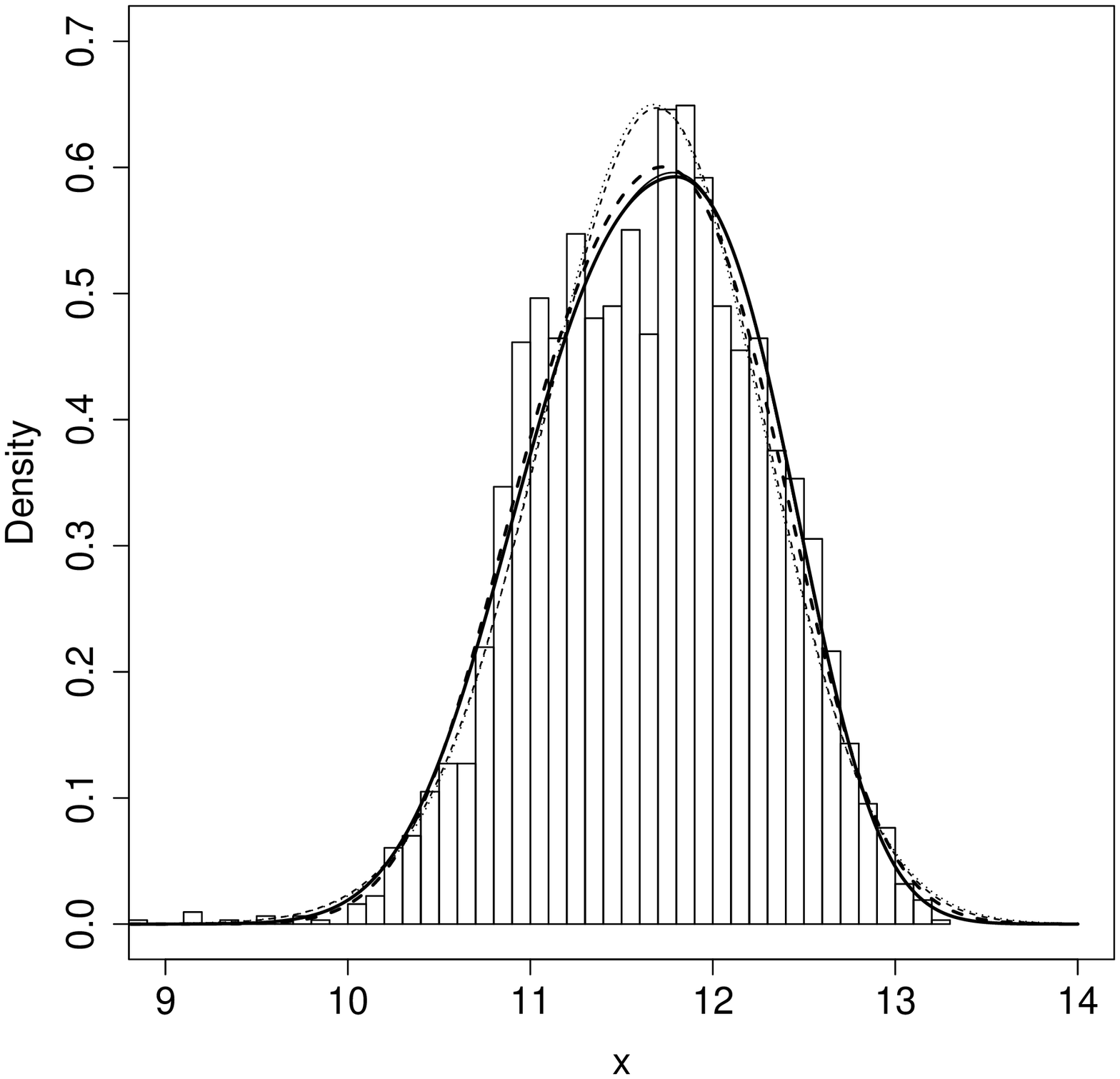,  height=6cm} &
\psfig{figure=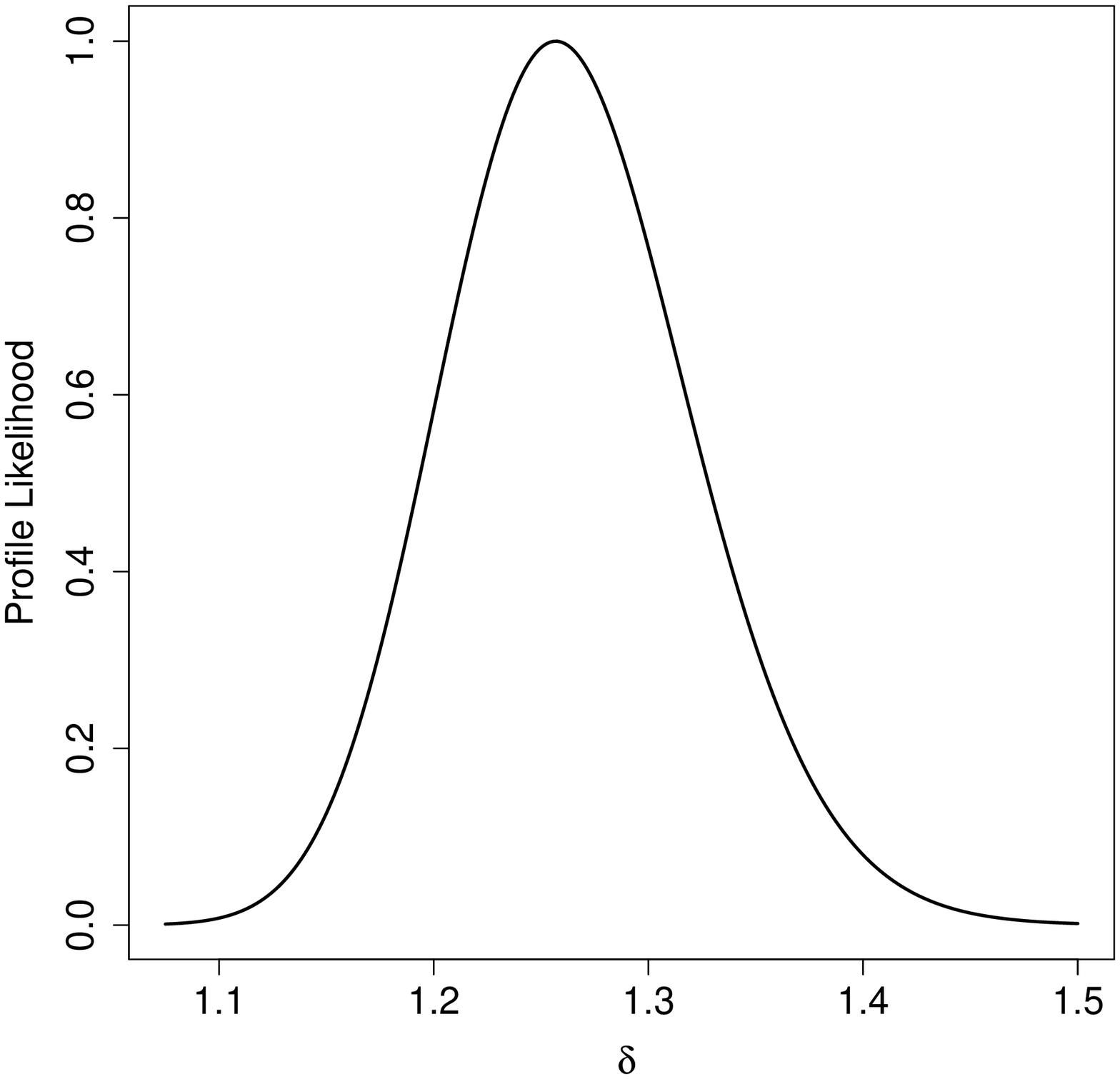,  height=6cm}  \\
(a) & (b) \\
\psfig{figure=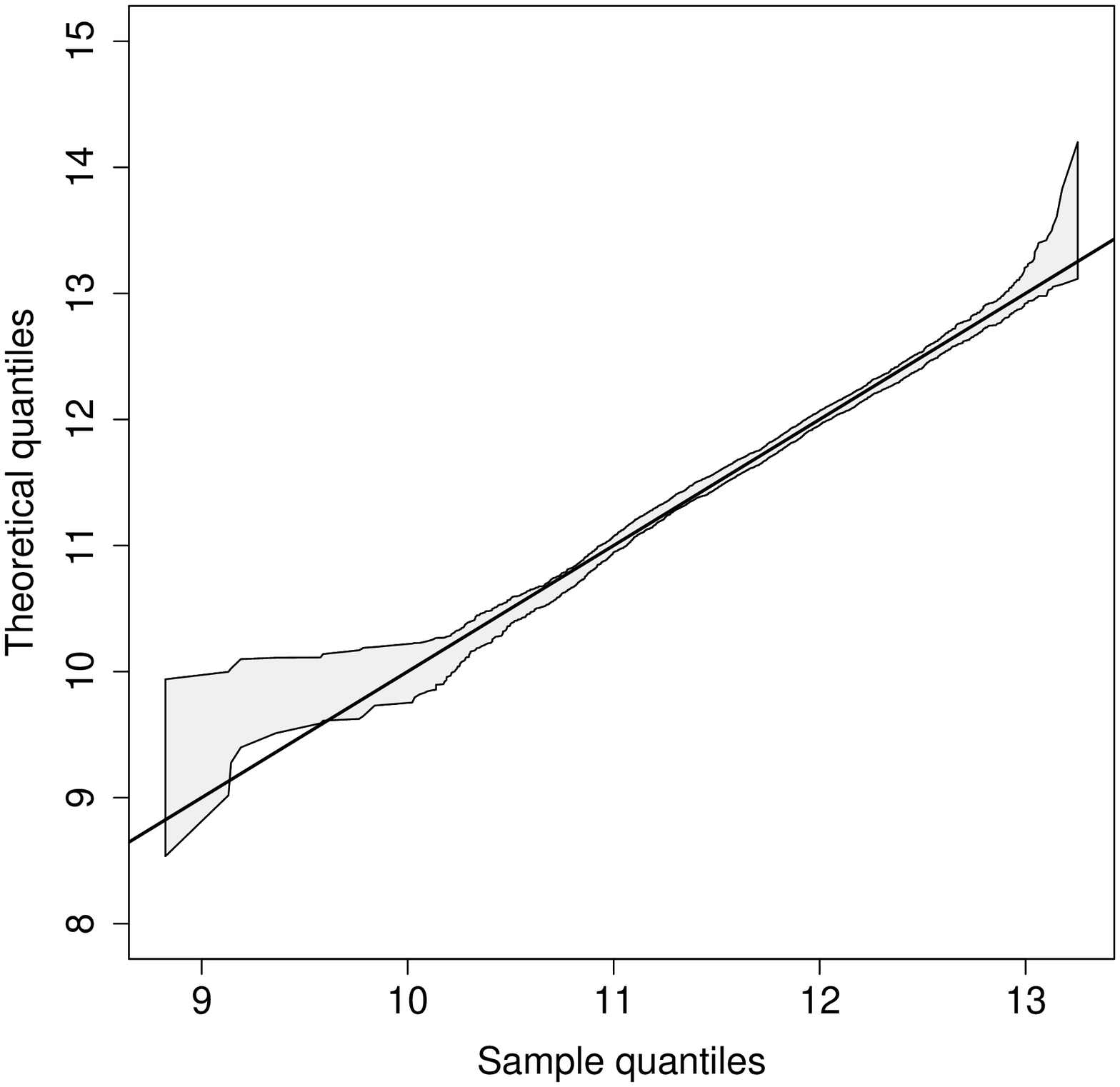,  height=6cm}&
\psfig{figure=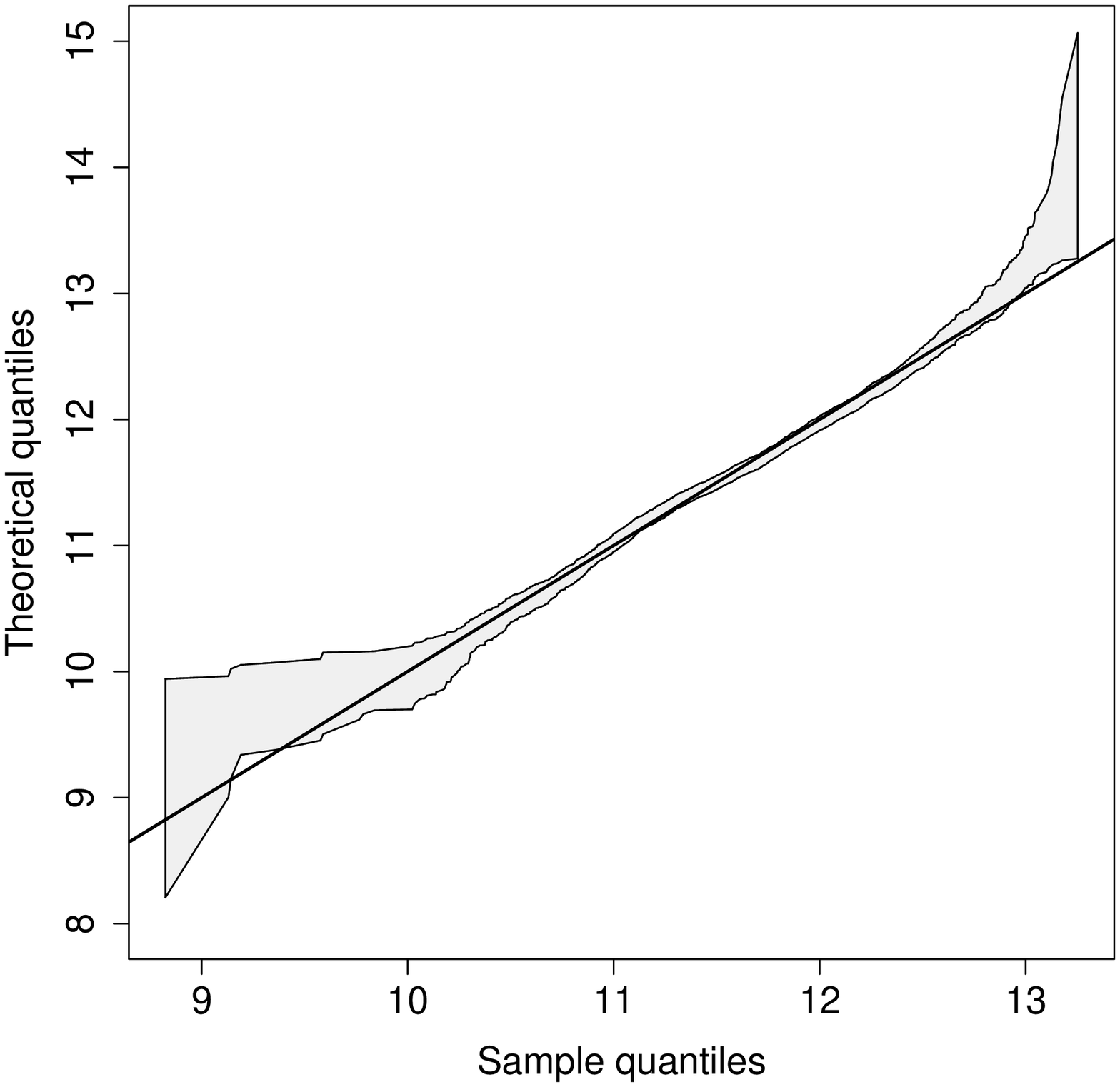,  height=6cm}\\
(c) & (d)
\end{tabular}
\end{center}
\caption{\small Internet traffic data: (a) Fitted densities: TP SAS (bold line), SS SAS (dashed bold line), SAS (solid line), NL (dashed line), ST (dotted line); (b) Profile likelihood of $\delta$ for the TP SAS model; (c) Envelope QQ-plots for the TP SAS model; (d) Envelope QQ-plots for the Normal model.}
\label{fig:FitInternet}
\end{figure}

The teletraffic data set studied in \cite{R10} contains $n=3143$ observations, which represent the measured transferred bytes/sec within consecutive seconds. \cite{R10} propose the use of a Normal-Laplace (NL) distribution to model these data after a logarithmic transformation. This model is the convolution of a Normal distribution and a two--piece Laplace distribution with location $0$, which is typically parameterised in terms of two parameters $(\alpha,\beta)$ that jointly control the scale and the skewness. The NL distribution has tails heavier than those of the normal distribution \citep{RJ04}. We compare the fit of the NL against the TP SAS and the SS SAS distributions, as well as some other competitors. The corresponding estimators and model comparison are presented in Table \ref{table:Internet}. We first observe that the SAS, TP SAS, and the SS SAS models suggest that the data presents lighter tails than normal, a feature that cannot be captured by the other competitors, including the NL model. An approximate 95\% confidence interval of $\delta$ in the TP SAS model (obtained as the $0.147$ profile likelihood interval) is $(1.15,1.38)$, which emphasises the need for a model that can capture lighter tails than normal. Moreover, the AIC and BIC largely favour the models with lighter tails than normal. Figure \ref{fig:FitInternet}a shows some fitted densities with the histogram of the data, and Figure \ref{fig:FitInternet}b shows envelope QQ-plots for the fitted TP SAS model. This graphical goodness of fit tool is obtained by generating $N=10,000$ samples of size $n=3143$ (same size as the original data) from the fitted TP SAS distribution and creating $N$ QQ-plots for each simulated sample against the original data. Using these $N$ QQ-plots, we can generate an envelope, by taking the minimum and maximum values of the QQ-plots at each quantile point, which is shown in the shaded area. This envelope is compared against a straight line with intercept 0 and slope 1, which represents a perfect fit. From Figure \ref{fig:FitInternet}c we can observe that, although the TP SAS beats the other competitors, the fit in the left tail is not entirely satisfactory. Figure \ref{fig:FitInternet}d shows that the normal model produces a poor fit on both tails. In fact, in the latest version of \cite{RS14b} it is shown that a more flexible (five-parameter) model is necessary to fit this data set adequately.

\begin{table}[ht]
\begin{center}
\begin{tabular}[h]{|c|c|c|c|c|c|c|}
\hline
Model & $\widehat{\mu}$ & $\widehat{\sigma}$ & $\widehat{\gamma}$ & $\widehat{\delta}$ & AIC & BIC\\
\hline
TP SAS & 11.80 & 0.85 & 0.14(0.08,0.20) &  1.26(1.15,1.38) & {\bf  5884.95} & {\bf 5909.16}\\
SS SAS & 11.53 & 0.85 & 0.26(0.04,0.52) & 1.24(1.14,1.36)  & 5900.26 & 5924.47 \\
SAS & 11.78 & 0.84 & -0.16(-0.24,-0.08) & 1.26(1.16,1.38)  & 5886.84 & 5911.05\\
NL & 11.78 & 0.56 & ($\widehat{\alpha}$) 8.39(6.20,9.04) & ($\widehat{\beta}$) 4.09(3.44,6.82)  &  5922.73 & 5946.94\\
skew--$t$ & 12.07 & 0.75 & -0.98(-1.36,-0.64) & 1057.37(108.39,6531.52)  &  5919.52 & 5943.73\\
SN & 12.09 & 0.76 & -1.04(-1.34,-0.67) & --  &  5917.04 & 5935.20\\
Normal & 11.65 & 0.62 & -- & --  &  5925.37 & 5937.47\\
\hline
\end{tabular}
\caption{\small Internet data: Estimation and model comparison (95\% likelihood-confidence intervals are in brackets).}
\label{table:Internet}
\end{center}
\end{table}

\section{Concluding Remarks}

We have introduced and studied the two--piece sinh--arcsinh (TP SAS) distribution, which contains the normal distribution
as well as symmetric and asymmetric models with varying tail--weight. The distribution was derived by applying the two--piece transformation to the symmetric sinh--arcsinh distribution (SAS) proposed by \cite{JP09}. Unlike the SAS distribution, the TP SAS distribution can produce models that cover the whole range of some common measures of skewness, and we have shown that its shape parameters have interpretable separate roles. The performance of the proposed distribution was illustrated using a publicly available data set. We have developed the `TPSAS' R package, where we implement the density function, distribution function, quantile function, and random number generation for the TP SAS model. We have also emphasised the need for conducting an integral model selection in which both a model selection tool and the inferential properties of the models in question are taken into consideration. As noted  by \cite{CVM13}, it is sensible to decide on the distribution to be used in modelling data on the basis of interpretation of its parameters rather than only the best fit, especially when competitor models produce a similar fit. A similar discussion, although in a more general context, was recently presented by \cite{J14}. We recommend the use of the profile likelihood for the construction of confidence intervals for parameters, rather than standard deviations based on asymptotic normality, given that the likelihood function is typically asymmetric for moderate sample sizes. Hence, the use of standard errors would lead to confidence intervals with the wrong coverage.

We conclude by pointing out other contexts where the proposed models can be of interest. \cite{WD10} employ a Generalised Extreme Value distribution as a link function in binary regression. They mention that it would be desirable to use ``a distribution such that one parameter would purely serve as skewness parameter while the other could purely control the heaviness of the tails'': we have shown that the TP SAS distribution has this property. In addition, the TP SAS link avoids a problem pointed out by \cite{J13} with their proposed flexible link: ``One potential problem with the proposed power link is that the power parameter $r$ influences both the skewness and the mode of the link function pdf'': for the TP SAS distribution, the parameter $\mu$ controls the mode, while $\gamma$ controls the mass cumulated on each side of the mode of the density. However, one has to be careful when using links with skewness and kurtosis parameters, as binary data typically carry little information about the tails of the link. Another potential use of the TP SAS distribution is to model the residual errors in a linear regression model. Linear regression models with parametric flexible errors have been mainly studied using the skew-$t$ distribution \citep{AG08}.


\end{document}